\documentclass{sig-alternate}
\usepackage{url}

\makeatletter
\def\@copyrightspace{\relax}
\makeatother

\begin{document}
%
\conferenceinfo{MM'16}{, Oct 15-Oct 19, 2016, Amsterdam, Netherlands.}

\title{Leveraging Contextual Cues for Generating\\Basketball Highlights}

%
%
%
%
%

\numberofauthors{3} 
%
\author{
%
%
\alignauthor
Vinay Bettadapura$^{\text{1}}$\\
       \email{\small vinaykb@google.com}
    \and
    Caroline Pantofaru$^{\text{1}}$\\
	\texttt{\small cpantofaru@google.com}
	\and
	Irfan Essa$^{\text{1},\text{2}}$\\
    \texttt{\small irfan@cc.gatech.edu}
	\and
    {\small $^{\text{1}}$Google Inc., Mountain View, CA, USA}
      \\
    {\small $^{\text{2}}$Georgia Institute of Technology, Atlanta, GA, USA}
    \\
        \url{http://www.vbettadapura.com/highlights/basketball}
}

\maketitle
\begin{abstract}

The massive growth of sports videos has resulted in a need for automatic generation of sports highlights that are comparable in quality to the hand-edited highlights produced by broadcasters such as ESPN. Unlike previous works that mostly use audio-visual cues derived from the video, we propose an approach that additionally leverages contextual cues derived from the environment that the game is being played in. The contextual cues provide information about the excitement levels in the game, which can be ranked and selected to automatically produce high-quality basketball highlights. We introduce a new dataset of 25 NCAA games along with their play-by-play stats and the ground-truth excitement data for each basket. We explore the informativeness of five different cues derived from the video and from the environment through user studies. Our experiments show that for our study participants, the highlights produced by our system are comparable to the ones produced by ESPN for the same games.

\end{abstract}

\category{H.3.1}{Information Storage and Retrieval}{Content Analysis and Indexing}[Indexing methods]
\category{I.2.10}{Artificial Intelligence}{Vision and Scene Understanding}[Video analysis]

\terms{Algorithm, Experimentation, Measurement.}

\keywords{Basketball Video, Video Highlights, Content Analysis, Highlight Ranking.}

\section{Introduction}

In recent years, there has been a proliferation in sports broadcasting due to the large number of games played across college and professional leagues throughout the year. Due to time limitations and regional differences, sports fans have a hard time watching the games live and keeping up with their favorite teams and players. This necessitates the need for high-quality sports highlights that allow the viewers to watch the interesting and exciting moments of the games at their own convenience. However, hand-editing the videos to generate the highlights is a time-consuming process and is not scalable, especially when we want to generate highlights of various lengths for all the different games depending on the needs of the viewers. 

Previously published works that automatically produce sports highlights have mostly focused on audio-visual cues that can be derived from the videos. However, the environment within which the game is taking place provides us with rich contextual information that can be leveraged to produce better quality highlights. The players are active within the environment and the audience reacts to their actions with a range of emotions ranging from excitement to frustration. Sensors such as cameras and microphones are setup by the broadcasters which capture the player activity and the audience reaction (both audio and video). There are also several expert third party observers within the environment such as referees, coaches, commentators, and on-court statisticians. The data from these observers coupled with the video data from the broadcast videos provides rich contextual cues that can be leveraged to better understand the sporting scene.

In this paper, we focus on basketball, which is the third most popular sport in the US (after Football and Baseball) \cite{GallupSports}. Basketball games are typically held in the indoor stadiums and gymnasiums of schools and colleges and provides us with a representative test-bed to develop our methodology and evaluate it. However, the system presented here is general and can be applied to any sport. Generating the highlights for an entire sports game involves understanding the salient moments of the game, generating an excitement-based rank-ordering of the plays, segmenting and extracting them from the broadcast video, and selecting the top clips to generate the game highlights. In the context of this study, we define a sports highlight as a ``highlight reel" that showcases the top $n$ exciting moments of the game in a chronological order.

The contextual cues used in our approach are derived from two sources within the basketball environment: (1) microphones that capture the audience and commentator audio, and (2) the play-by-play stats data from the on-court statisticians. From these two environmental sources, we extract four different cues: ``Audio", "Score Differential", ``Player Ranking" and ``Basket Type". Finally, a fifth cue ``Motion" is extracted from the broadcast video which captures the magnitude of player and camera motion. For each basket within a given game, the data from these five cues is combined to generate an ``excitement score" for the basket. Once all the baskets have been scored, we can then rank them by their excitement scores and pick the top $n$ exciting clips and use them to generate the game highlights.

In order to conduct this study, we built a database of 25 NCAA games (played between February and March of 2015) totaling 35.44 hours of basketball footage along with the corresponding play-by-play stats data. There are a total of 1,173 baskets across these 25 games. We conducted extensive user-studies using Amazon's Mechanical Turk in order to obtain ground-truth on the excitement levels for each of these 1,173 baskets. The ground-truth data was then used to study the effectiveness of each of the cues as an indicator of how exciting a basket is. Finally, the five cues are combined using a weighted sum wherein the weights are learned from the data using 25-fold cross-validation (where we train using 24 games and test on the held-out game and repeat). We conducted a second round of user-studies for evaluation purposes, wherein we show (1) the effectiveness of cue-combination over each of the individual cues, and (2) that the highlights that we generate with our cue-combination are comparable to the highlights produced by ESPN for those games, wherein the ESPN highlights were regenerated using the same video pipeline that was used to generate our highlights (in order to make our highlights and ESPN highlights look visually similar in the user studies).

\textbf{Contributions:} Our contributions are as follows: (1) We present a method to leverage contextual cues from the environment to understand the excitement levels within a basketball game and automatically produce basketball highlights, (2) We introduce a new dataset of 25 NCAA games (35.44 hours of video with 1,173 baskets) along with the play-by-play stats and the ground-truth excitement data for each basket (we will make this dataset public to the research community), (3) We explore five different cues and study their effectiveness in determining the excitement of baskets through an extensive user study, and (4) We conduct user studies and show that the final highlights that we produce are comparable to the ESPN-ranked highlights.

\section{Related Work}

Sports analytics and summarization has been an active area of research for the past two decades. Most of the work has been on analyzing broadcast videos from sports such as soccer, basketball, hockey, football and tennis. Professional broadcast videos (such as videos from ESPN) contain replays which can be extracted by detecting the logo-sweeps (shown before and after the replays) and the ``arousal" level of the replays can be computed using the audience's audio energy and the amount of camera motion in order to rank the replays in terms of their excitement level \cite{zhao2006highlight}. Slow-motion replays can also be detected using Hidden Markov Models (HMMs) and Support Vector Regressors \cite{tong2005highlight} and summaries can be generated by concatenating the detected replays. When replays are not available in the broadcast video, baskets can be detected by detecting breaks in the game and using object detectors to detect the referee and the penalty boxes to make informed choices about the importance of different plays during the games \cite{ekin2003automatic}. However, these approaches are limiting since detecting replays and slow motions and using those clips in the highlights will give us a highlight reel that has only those baskets for which replays or slow motions were shown. There could be many other exciting baskets that are missed because the broadcast director chose not to show the replays or slow motions for those baskets.

Audio plays a crucial role in detecting highlights in sports. The energy of the crowd and the excitement in the commentator's voice provides useful cues that can be used to pick exciting moments in the game. Audio-based architectures for sports summarization have been developed that extract audio features and classify the audio segments as applause, cheering, music, speech, etc. and also perform background noise modeling to further refine the results \cite{xiong2003audio}. Along with audio, the amount of motion within the broadcast footage also helps identify exciting moments. The motion content of videos is encoded into the MPEG-7 motion activity descriptors. These motion vectors can be quantized and combined with the audio features to generate cumulative rankings of exciting moments \cite{xiong2003generation,liu2009framework}. Audio and motion curves can also be combined to generate excitement time curves wherein the maximas represent the game highlights \cite{hanjalic2003generic, hanjalic2003multimodal} and the minimas around the maximas can be used to determine the segment boundaries of the highlight clips \cite{mendi2013sports}. Motivated by these approaches, we investigate the use of audio and motion in our system.

An interesting area of research in sports summarization involves studying the problem from a affective rather than a cognitive point-of-view. The cognitive point-of-view is fact-based, wherein the features used for highlight detection are facts such as audio energy, amount of motion, position of ball, etc. In contrast, the affective point-of-view is emotion-based and tries to understand the human emotions within the game. Affect has three underlying dimensions: valence (ranging from pleasant to unpleasant), arousal (ranging from excited to peaceful) and control (no-control to full-control). All of human emotions can be mapped in as a set of points in this 3D VAC space. Computational methods have been developed to compute the valence and arousal using video and audio features and using them to find highlights in both sports and movies, thereby generating summaries from an affective point-of-view \cite{hanjalic2005affective}.

In the past few years, with the proliferation of social media and blogging websites, researchers have turned their attention to ``crowd-sourced" sports summarization techniques. People watching broadcast games use Twitter to tweet their reactions. Mining the Twitter data for relevant tweets and looking for times when there is a spike in the volume of tweets gives us the moments in time that the crowd deems to be interesting \cite{nichols2012summarizing,hannon2011personalized}. Crowd-sourced summaries have several differences over traditional summaries generated by sports professionals. In crowd-sourced summaries the highlights that get selected include interesting plays that require high degree of skill (expected and easy plays are ignored), controversial plays and unusual occurrences (like fights and stunts) and ``lowlights" which are moments in time when the fans are frustrated and angry at their favorite teams \cite{tang2012epicplay}. Other methods include analyzing web-casting text and social media blogs, aligning them with the broadcast videos and looking for highlights using player popularity and crowd sentiments \cite{tjondronegoro2011multi,xu2008novel}.

While most of these published works use cues derived only from the video data, the only other environmental contextual cue that is used is the commentator and audience audio. In the proposed work, we look at the play-by-play stats that is obtained from the on-court statisticians, a source of data that has largely been ignored by the research community. We show that the play-by-play stats contain a wealth of information that can be leveraged to generate highlights that are comparable to the ESPN-ranked highlights.

\section{Methodology}

\begin{figure} 
\begin{centering}
\includegraphics[width=1.0\columnwidth]{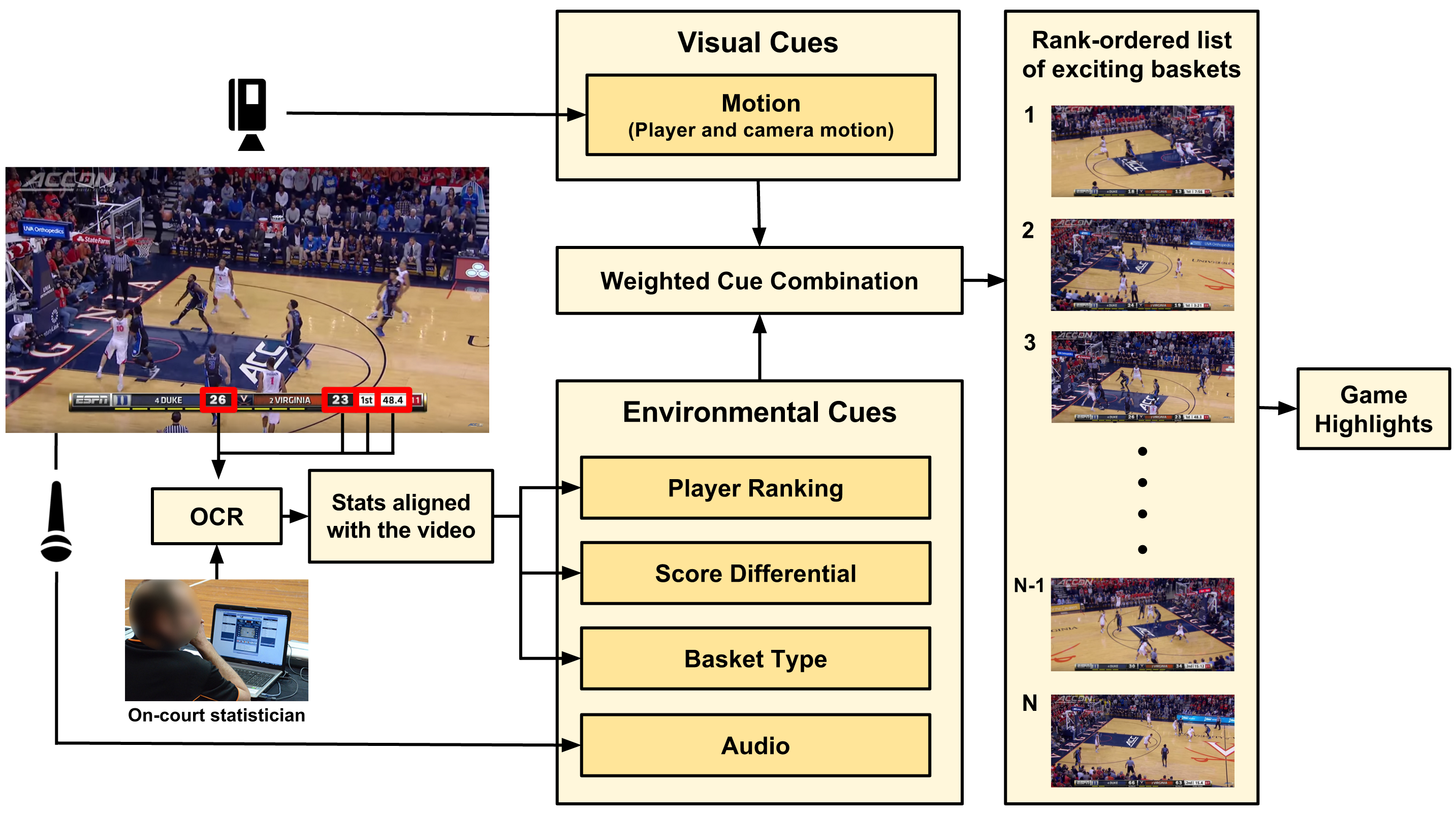}
\par\end{centering}
\caption[Basketball highlights system overview]{\label{fig:sports-summarization}An overview of our system that uses visual and environmental contextual cues for automatically producing basketball highlights.}
\end{figure}

An overview of our system is given in figure \ref{fig:sports-summarization}. The five different cues (four environmental and one visual) form the core component of our system. Let us look at each of these five cues in detail:

\subsection{Cue 1: Audio}

Gymnasiums and stadiums are equipped with microphones that capture the commentator and audience audio. Exciting baskets typically draw loud cheers from the audience and result in an elevation in the loudness and pitch in the commentator's voice. The changes in their audio levels are important contextual cues that are indicative of how exciting a basket is \cite{hanjalic2003generic, hanjalic2003multimodal, mendi2013sports, xiong2003audio}.

\begin{figure}
\begin{centering}
\includegraphics[width=1.0\columnwidth]{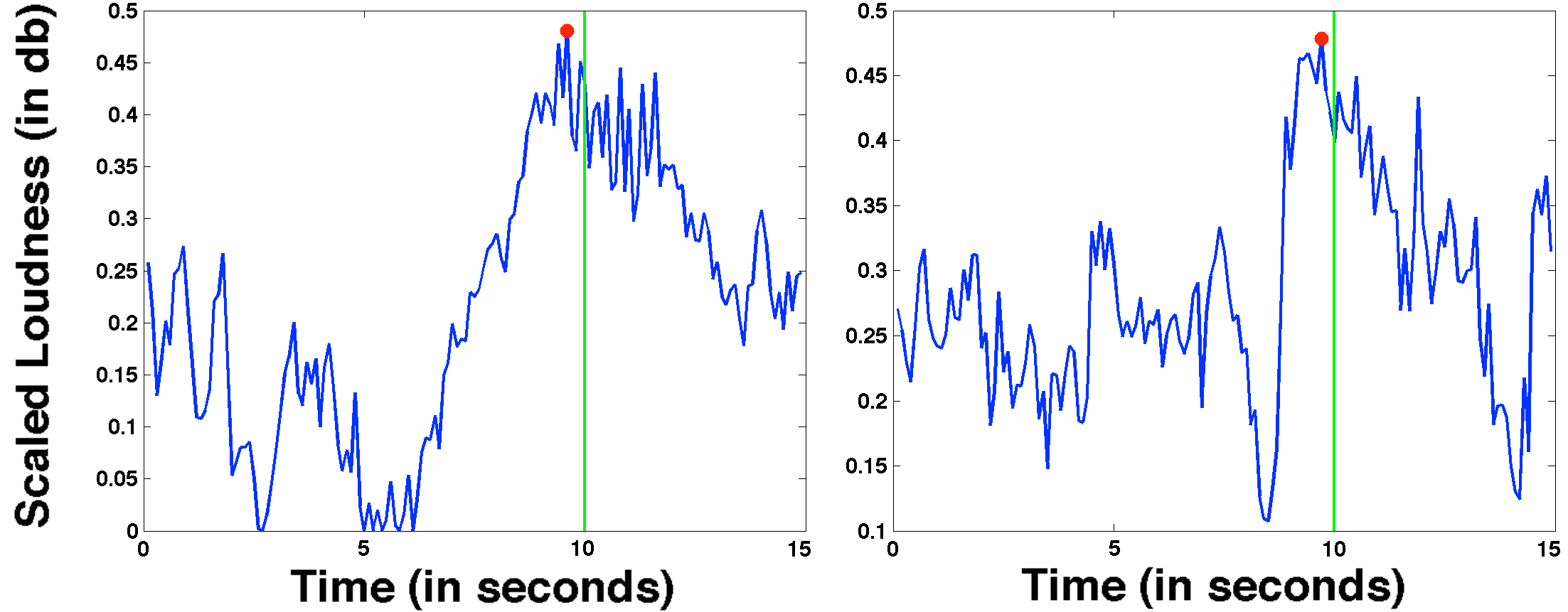}
\par\end{centering}
\caption[Audio loudness plots for two sample baskets]{\label{fig:audio-plots}Audio loudness plots for two sample baskets. The red dot represents the time when the basket was scored and the green line (at the 10 second mark) represents the time when the scoreboard was updated to reflect the new scores. We can see that (1) the audio excitement peaks when the basket occurs, and (2) there is a slight delay of a few milliseconds between when the basket occurs and the scoreboard updates (the time difference between the red dot and the green line). The audio excitement drops soon after as the game continues.}
\end{figure}

In our study, the audience and commentator audio are obtained from the broadcast video and thus unavailable on two separate channels for analysis. Let us denote this signal as $a$. Before we can compute statistics on $a$, it has to be pre-processed in order to obtain the true audio loudness, $a_l$,  based on human perception of loudness. We perform this pre-processing by following the audio filtering guidelines provided by the International Telecommunications Union (ITU) \cite{itu2011itu}. The first stage of pre-processing applies a pre-filtering of the audio signal prior to the Leq(RLB) measure. The pre-filtering accounts for the acoustic effects of the head, where the head is modeled as a rigid sphere. The second stage of the algorithm applies the RLB weighting curve, which consists of a single high-pass filter. With the pre-filter and the RLB filtering applied, the mean square energy in the measurement interval $T$ is then measured. Once the weighted mean square level has been computed for each channel, the final step is to sum the $N$ audio channels. The audio loudness levels obtained using this approach is shown for two sample baskets in figure \ref{fig:audio-plots}.

The true audio level, $a_l$, obtained using the ITU's guidelines has been shown to be effective for use on audio programs \emph{that are typical of broadcast content} which makes it the ideal audio pre-processing step for our system. Once the audio signal has been pre-processed, the measure of excitement for a given basket $b$ is computed as
\begin{equation}
A_b=\sum_{i=1}^{m}{p_i(a_l)},
\end{equation}
where $p_i(a_l)$ is the $i^{th}$ highest loudness peak in a 4 second window around the basket (3 seconds before the basket and 1 second after the basket). The overall audio loudness level for each basket, $A_b$, is obtained by summing the top $m$ peaks. Empirically, for our NCAA dataset, $m$ was determined to be 7. Finally, for each game, the $A_b$ values for all the baskets are normalized between 0 and 1 by computing the $min$ and $max$ values across all baskets for that game.

\subsection{OCR: Aligning the Stats With the Videos}

A main source of environmental context in our approach, is the play-by-play stats data that is generated by the on-court statisticians. The play-by-play data can be available in near-real-time or it can be available post-game. In either case, the stats need to be aligned with the broadcast video in order to determine when the particular play mentioned in the play-by-play stats actually occurred in the video. Unfortunately, the play-by-play stats are specific to a game and not specific to any particular broadcast video of the game. Hence they do not contain the video time-stamp of when the play occurred in the broadcast video.

\begin{figure}
\begin{centering}
\includegraphics[width=1.0\columnwidth]{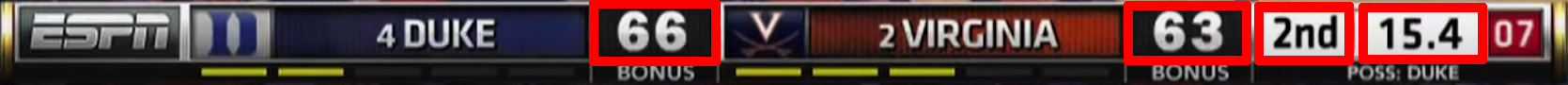}
\par\end{centering}
\caption[Basketball graphics overlay]{\label{fig:graphics-overlay}A typical graphics overlay shown in basketball broadcasts. They contain four key pieces of information (shown highlighted using red boxes): (1) home team score, (2) visiting team score, (3) game period, and (4) game clock.}
\end{figure}

In order to align the stats with the video, we introduce a novel Optical Character Recognition (OCR) based technique. The broadcast videos have a graphics overlay which contain four key pieces of information: (1) home team score, (2) visiting team score, (3) game period, and (4) game clock. An example is shown in Figure \ref{fig:graphics-overlay}. Using the Tesseract OCR system \cite{smith2007overview}, these four values are read for each frame of the video and are stored along with the corresponding video time-stamp. Next, we parse the play-by-play stats file and match the stored OCR info with each of the stats. This results in a mapping of the stats to the broadcast video. \emph{Example:} Say for a particular game, for a particular frame of the video, using OCR on the graphics overlay, we know that the home team score changed from ``35" to ``38" while the visiting team score was ``29" during the ``1st half" of the game at game clock ``12:22" and when the video timestamp was ``28:34". While parsing the play-by-play stats, we see an entry ``Player: Jahlil Okafor, Basket Type: 3-Pt Jump Shot, Game Period: 1st Half, Home Score: 38, Visiting Score: 29, Game Clock: 12:22". By matching this entry with the OCR data, we can see that this particular 3-Pt Jump Shot basket by Jahlil Okafor took place at time-stamp ``28:34" in the video. This allows us to align the rich contextual info from the stats with the corresponding basket within the video.

Next, we extract three different cues from the stats data that are indicative of the excitement levels within the game: ``Player Ranking", ``Score Differential" and ``Basket Type" . Each of these cues are described below.

\subsection{Cue 2: Player Ranking}

Baskets by ``star" (top-ranked) players tend to be generate more excitement among basketball fans than baskets by other lower ranked players. Also, our analysis of ESPN highlights of 10 NCAA games showed that ESPN tends to favor baskets by the star athletes and showcases them more in the highlights. The game stats may or may not contain the player ranking, but they almost always have the data on each player's Points-Per-Game average (PPG). PPG has very strong correlation with player ranking and can be used as a proxy when player ranking data is not available. For each game, we normalize each player's PPG between the $min$ and $max$ PPG of all the players in that game (across both the teams). For each basket $b$, the scaled PPG value, $P_b$, of the player who made the basket gives us the ``player ranking" excitement score for that basket.

\begin{figure}
\begin{centering}
\includegraphics[width=1.0\columnwidth]{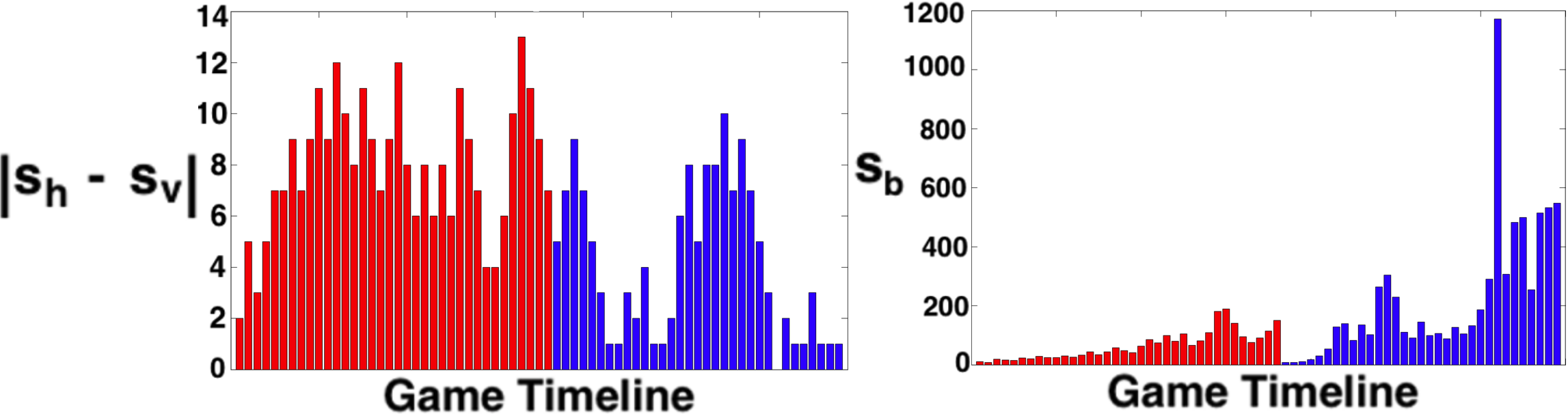}
\par\end{centering}
\caption[Score differential plots for a sample game]{\label{fig:score-diff}Score differential plots for a sample game. The $x$-axis represents the game timeline and the red and blue sections of the plots represent the 1\textsuperscript{st} half and 2\textsuperscript{nd} half of the game respectively. \textbf{Left:} $|s_h-s_v|$, the absolute score differential. \textbf{Right:} $S_b$, the inverted score differential weighted by the game-clock for each period of the game (see equation \ref{eq:score-diff}).}
\end{figure}

\subsection{Cue 3: Score Differential}

People tend to find a game to be more exciting when the game is close (``neck-to-neck") and less exciting when one team has a huge lead over the other. Furthermore, the game tends to be more exciting if the scores of the two teams are close towards the end of the game period. For a given basket $b$, if the home team score is $s_h$ and the visiting team score is $s_v$, then the ``score differential" excitement, $S_b$, for that basket is computed as
\begin{equation} \label{eq:score-diff}
S_b=\frac{1}{\left ( \left | s_h-s_v \right | +1\right)}*(1200-g_s),
\end{equation}
where $g_s$ is the game clock in seconds. As the score differential gets smaller, the excitement score $S_b$ increases. Each game period is 20 minutes long and the game clock counts down from 20:00 (1200 seconds) to 00:00 (0 seconds). So, the score differential is weighed by the amount of time remaining in the game period. Lower score differentials towards the end of the game period will get higher weights, and in-turn, higher excitement scores. The score differential plots for a sample game are shown in figure \ref{fig:score-diff}. The first half of the game is shown in red and the second half of the game is shown in blue. We can see the absolute score differential on the left and our ``score differential" excitement, $S_b$, on the right. As with the other cues, the $S_b$ scores are normalized between 0 and 1 for each game.

\begin{figure}
\begin{centering}
\includegraphics[width=1.0\columnwidth]{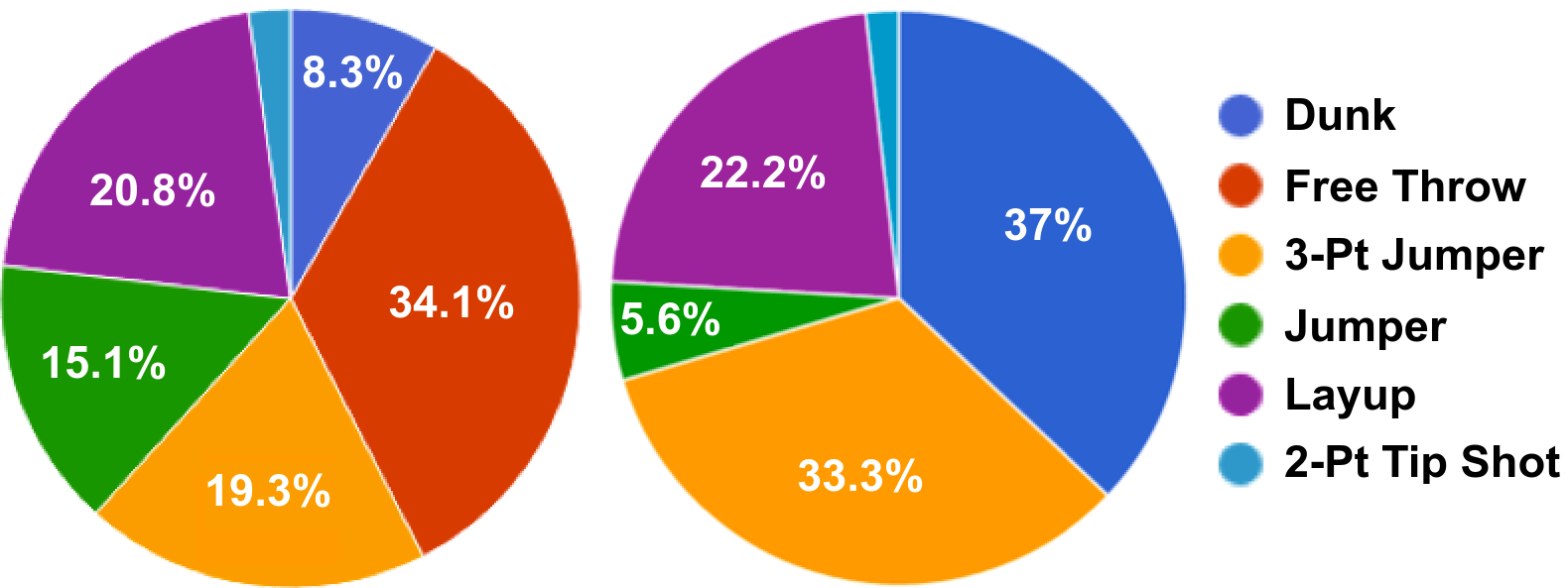}
\par\end{centering}
\caption[Preference in baskets shown in the highlights based on basket type]{\label{fig:basket-type}This figure shows the preference in baskets shown in the highlights generated by ESPN based on the basket type. \textbf{Left:} The distribution of baskets based on basket type across 10 full-length NCAA games. \textbf{Right:} The distribution of baskets based on basket type across the 10 highlights produced by ESPN for the same 10 NCAA games.}
\end{figure}

\subsection{Cue 4: Basket Type}

There are five types of baskets that are shown in basketball highlights: ``Dunk", ``Jumper", ``Layup", ``Two Point Tip Shot", and ``Three Point Jumper" (``Free Throws" are typically not featured in the highlights). Each of these five baskets require different techniques and skills. Basketball fans tend to find some basket types more exciting than others. For example, the dunk shot is universally considered to be one of the most exciting basketball plays and is prominently featured in the highlights produced by ESPN. This is illustrated in figure \ref{fig:basket-type}. On the left, we can see the distribution of baskets based on the basket type across 10 full-length NCAA games. On the right, we can see the distribution of baskets based on the basket type across the 10 highlights produced by ESPN for the same 10 NCAA games. We can clearly see that ESPN favors certain basket type over others. Although ``Free Throw" occurs 34.1\% of the time, they are almost never featured in the highlights due to the fact that a ``Free Throw" is not very exciting to watch. However, ``Dunk" occurs only 8.3\% of the time, but is featured in 37\% of the highlights. This is due to the fact that the viewers love watching a ``Dunk" and consider it to be much more exciting than the other basket types.

These 5 baskets can be rank-ordered in 5! = 120 different ways. Each of these 120 different basket rankings were evaluated on our NCAA dataset and the ranking that best matched the user-generated ground-truth was chosen. The ranking with the best match was: ``Dunk" \textgreater ``Two Point Tip Shot" \textgreater ``Three Point Jumper" \textgreater ``Layup" \textgreater ``Jumper". Using this ranking, for each basket $b$, the corresponding basket type's rank position, $B_b$, gives us the ``basket type" excitement score for that basket. The $B_b$ scores are normalized between 0 and 1 for each game.

\subsection{Cue 5: Motion}

The amount of player motion during a given play is usually an indication of how exciting the play is. For example, a ``Free Throw" which has minimal player motion is less exciting than a ``Dunk" wherein all the players are in rapid motion. The amount of camera motion is also indicative of the excitement levels of the game. For example, a large panning motion is involved when something exciting happens, such as a player running from one end of the court to another with the ball. In contrast, a free-throw has almost no camera movement and is less exciting than the other types of plays.

For each basket, we computed the optical flow using KLT tracking \cite{shi1994good} across all the frames. Camera motion was determined by computing the dominant optical flow and player motion was computed by subtracting the camera motion from the overall flow. For each basket $b$, the corresponding camera motion magnitude $M_b^c$, player motion magnitude $M_b^p$, and the overall motion $M_b$, gives us the ``motion" excitement scores for that basket. Our experiments showed that $M_b$ was a better indicator of excitement when compared against $M_b^c$ and $M_b^p$ individually. The $M_b$ scores are normalized between 0 and 1 for each game.

\subsection{Cue Combination}

Once all the 5 cues have been extracted for all the baskets and have been normalized for each game so that they have the same scale, we can combine them using a weighted sum. The final score, $C_b$, for each basket $b$ is given by
\begin{equation}
C_b=\omega_1*A_b+\omega_2*P_b+\omega_3*S_b+\omega_4*B_b+\omega_5*M_b,
\end{equation}
where $\sum_{i=1}^{5}w_i=1$. For our 25 game NCAA dataset, the weights are learned using 25-fold cross-validation. One of the games is held out as test and the weights are learned using the ground-truth excitement data from the other 24 games. The process is repeated 25 times, each time holding out a different game for testing. The final cues weights are computed by averaging the weights across all the 25 runs.

\subsection{Generating Highlights}

\begin{figure}
\begin{centering}
\includegraphics[width=1.0\columnwidth]{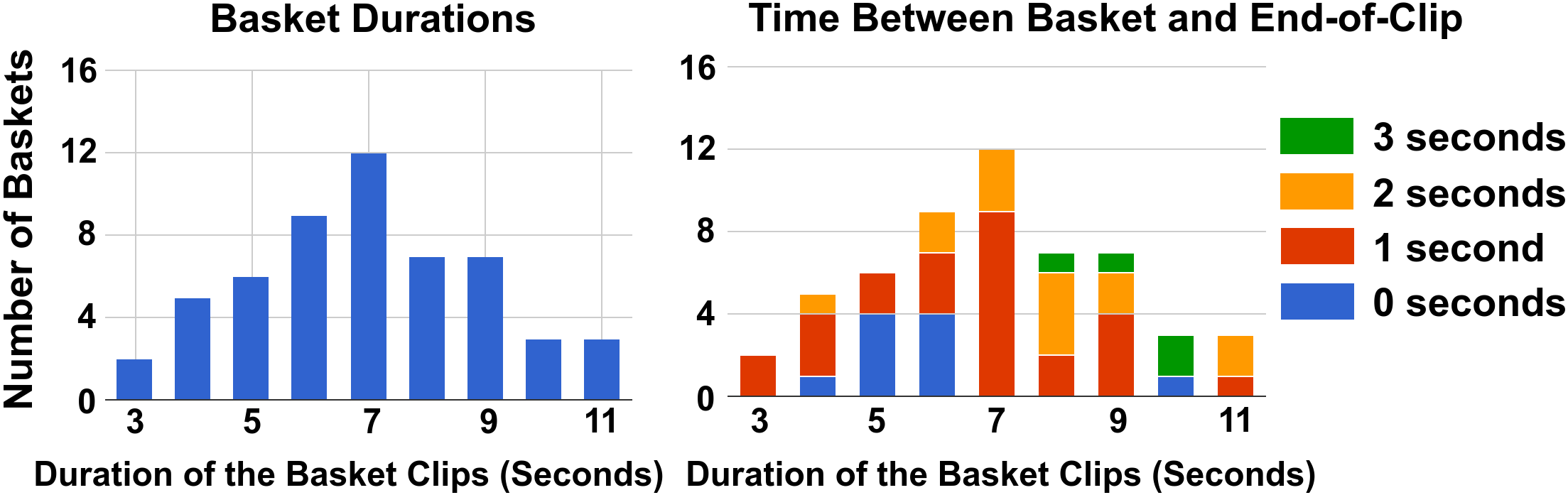}
\par\end{centering}
\caption[Histogram of duration of the baskets shown in the ESPN clips]{\label{fig:basket-durations}\textbf{Left:} Histogram of duration of the baskets shown across 10 ESPN highlights. We can see that ESPN prefers to show basket clips that are 6 to 7 seconds long. \textbf{Right:} Within each duration, we can see the time elapsed between when the basket occurred and the clip ended. For example, there are 12 baskets which were 7 seconds long. Out of these 12 baskets, 9 baskets had 1 second duration between when the basket happens and the clip ends and the other 3 baskets had 2 seconds duration.}
\end{figure}

With the final cue combination score for all the baskets of a game, we can rank-order them in terms of their excitement scores. The next step is to put them together to form the game highlights. The top $n$ exciting baskets are selected from the rank-ordered list, extracted from the broadcast video, sorted by time-stamp (so that the baskets appear in chronological order), and put together to form the game highlights. The value of $n$ depends on the length of the desired summary.

Each of the $n$ clips that were extracted from the broadcast video were 7 seconds long and had 1.5 second duration between when the basket occurred and the clip ended. These numbers were learned from the data by studying ESPN highlights. Figure \ref{fig:basket-durations} shows the histogram of durations of the baskets shown across 10 ESPN highlights and the time elapsed between when the basket occurred and the clip ended. We can see that ESPN prefers to have basket clips that are 6-7 seconds long with 1-2 seconds between the basket and the end of the clip.

\section{Evaluation}

In this section, we describe our NCAA dataset and the process by which we collected ground-truth pairwise excitement for each of the baskets. This is followed by the evaluation of each of our cues and a demonstration of the effectiveness of weighted cue combination as a predictor of excitement. Finally, we evaluate the highlights generated using cue combination against highlights generated using each individual cue and also compare against the highlights produced by ESPN.

\begin{table*}[]

\caption[Inter-rater reliability for pairwise excitement A/B tests]{\label{tab:inter-rater-table}The inter-rater reliability metrics for our A/B tests on assessing the pairwise excitement of the baskets in our NCAA dataset.}

\centering
\begin{tabular}{|r|r|r|r|r|l|}
\hline
\multicolumn{1}{|c|}{\textbf{\begin{tabular}[c]{@{}c@{}}Agreement\\ between\\ N or more\\ users\end{tabular}}} & \multicolumn{1}{c|}{\textbf{\begin{tabular}[c]{@{}c@{}}Number of\\ baskets\end{tabular}}} & \multicolumn{1}{c|}{\textbf{\begin{tabular}[c]{@{}c@{}}Average\\ pairwise\\ agreement\\ percentage\end{tabular}}} & \multicolumn{1}{c|}{\textbf{\begin{tabular}[c]{@{}c@{}}Average\\ pairwise\\ Cohen's\\ kappa\end{tabular}}} & \textbf{\begin{tabular}[c]{@{}l@{}}Fleiss'\\ kappa\end{tabular}} & \multicolumn{1}{c|}{\textbf{Interpretation}} \\
\hline
\hline 
8                                                                                                     & 1173                                                                             & 54.26\%                                                                                                  & 0.067                                                                                             & 0.067                                                   & Slight agreement                    \\ \hline
9                                                                                                     & 880                                                                              & 56.79\%                                                                                                  & 0.105                                                                                             & 0.105                                                   & Slight agreement                    \\ \hline
10                                                                                                    & 625                                                                              & 60.14\%                                                                                                  & 0.154                                                                                             & 0.154                                                   & Slight agreement                    \\ \hline
11                                                                                                    & 384                                                                              & 65.01\%                                                                                                  & 0.212                                                                                             & 0.213                                                   & Fair agreement                      \\ \hline
12                                                                                                    & 203                                                                              & 71.18\%                                                                                                  & 0.270                                                                                             & 0.270                                                   & Fair agreement                      \\ \hline
13                                                                                                    & 92                                                                               & 77.76\%                                                                                                  & 0.304                                                                                             & 0.305                                                   & Fair agreement                      \\ \hline
14                                                                                                    & 18                                                                               & 88.15\%                                                                                                  & 0.416                                                                                             & 0.382                                                   & Fair agreement                      \\ \hline
\end{tabular}
\end{table*}

\subsection{Dataset Description}

In order to gather ground-truth excitement data and evaluate our approach, we built a new basketball dataset. This dataset will be made public.

We collected 25 full-length broadcast videos of NCAA games from 2015 (March Madness) from YouTube. This is a total of 2,126.5 minutes (35.44 hours) of basketball videos. All the videos are 720p HD at 30fps.

For each of these games, we also collected the play-by-play stats data. Next, using the OCR technique described above, we aligned the play-by-play stats with the corresponding videos and extracted all the baskets such that each basket clip was 8 seconds long (5.5 seconds before the basket and 2.5 seconds after the basket). Free Throws were ignored since Free Throws are typically not shown in highlight reels (see Figure \ref{fig:basket-type}). This gave us a total of 1,173 baskets across all the 25 NCAA games with the corresponding stats data for each basket. The stats data contains the following information: (1) player name, (2) basket type, (3) home team score, (4) visiting team score, (5) game clock, and (6) game period.

For 10 out of the 25 games, we also have the game highlights produced by ESPN. The videos were collected from YouTube and are 720p HD at 30fps.

\subsection{Ground-Truth Pairwise Excitement}

Getting ground-truth excitement data on all the 1,173 baskets in our dataset lets us analyze each of our five cues individually and the effectiveness of cue combination. However, collecting this ground-truth is non-trivial. Users find it hard to subjectively rank a bunch of clips based on how exciting the clips are. The more the number of clips, the harder the task becomes. However, users find it fairly easy to pick the exciting clip given only two choices. Thus, in order to gather the ground-truth data, we conducted A/B test user studies on Amazon's Mechanical Turk where users were shown a basket from one of the games and another random basket from the same game and were asked the question ``Which of these two clips is more exciting to watch?". We took several steps to ensure the quality of these user studies:

\begin{figure}[t]
\begin{centering}
\includegraphics[width=1.0\columnwidth]{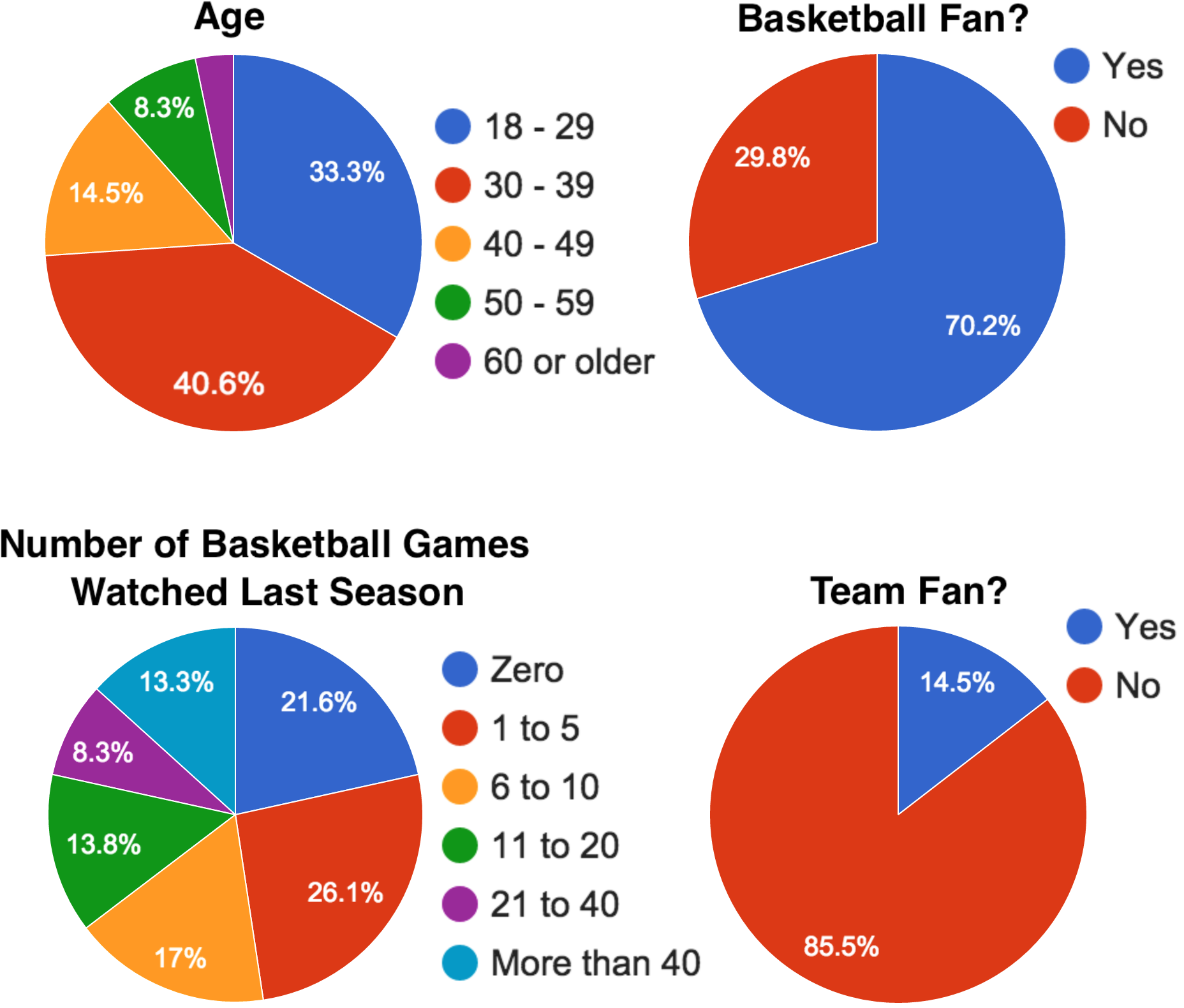}
\par\end{centering}
\caption[Ground-truth A/B testing: user distribution]{\label{fig:user_distribution}The distribution of 399 users across the different A/B tests for capturing the ground-truth pairwise excitement levels of the baskets. Top Left: Distribution across age groups. Top Right: Distribution on whether basketball fan or not. Bottom-Left: Distribution based on the number of games watched last season. Bottom Right: Distribution on whether team fan or not.}
\end{figure}

\begin{figure*}
\begin{centering}
\includegraphics[width=1.0\textwidth]{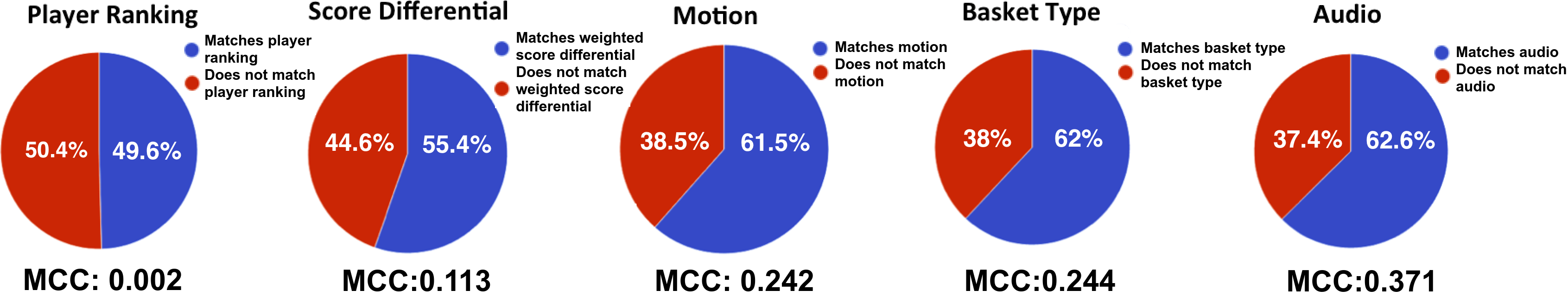}
\par\end{centering}
\caption[Performance of each individual cue]{\label{fig:cue-performance-fig}The performance of each cue as a predictor of the excitement levels of baskets.}
\end{figure*}

\begin{itemize}

\item Each A/B pair was shown to 15 different users in order to get good data and reduce the likelihood of selection based on chance. This resulted in 17,595 A/B tests (15 studies for each of the 1,173 baskets).

\item We required all Mechanical Turk users to have atleast 95\% approval rating and a minimum of atleast 1,000 previously approved tasks. The A/B tests took an average of 1 minute and 29 seconds and the users were paid \$0.03 per study.

\item The order in which the clips were shown in each A/B test was randomized. This further decreased the likelihood of user bias towards any one choice.

\item The users were asked additional questions as a part of each A/B test in order for us to gain more insight into who our users were. These additional questions were:

\begin{itemize}
\item ``Are you a basketball fan?" [options ``Yes, ``No"].
\item ``Are you a fan (or alumnus) of one of the teams in the clips?" [options ``Yes, ``No"].
\item ``How many basketball games did you watch last season?" [options: ``None", ``1 to 5", ``6 to 10",  ``11 to 20",  ``21 to 40", ``Greater than 40"].
\item ``What's your age?" [options: ``18 to 29", ``30 to 39", ``40 to 49", ``50 to 59", ``60 or older"].
\end{itemize}

\end{itemize}

After all the A/B test user studies were completed, we analyzed the data and found that there were 399 unique users who had participated in our studies. The distribution of the users based on the questions we asked is as shown in Figure \ref{fig:user_distribution}. We can see that the majority of our users are between the age range of 18 to 39, are mostly basketball fans, and mostly not fans of any of the two teams shown in the clips.

\subsection{Inter-Rater Reliability}

In order to compute the overall consensus across our 15 A/B tests for each of the 1,173 baskets, we compute two inter-rater reliability metrics -- the Fleiss' kappa and the average pairwise Cohen's kappa \cite{fleiss1973equivalence}. These statistical measures take into account the amount of agreement that could be expected to occur through chance. Table \ref{tab:inter-rater-table} shows the inter-rater reliability metrics for different values of $N$, where $N$ is the number of users who agreed that one basket was more exciting than the other. For example, from Table \ref{tab:inter-rater-table} we can see that there are 384 baskets for which 11 or more users (out of the total 15) agreed that one basket was more exciting than the other in the randomized A/B tests. This has a Fleiss' kappa of 0.213 which is interpreted as ``Fair Agreement".

\subsection{Evaluating Individual Cues}

The pairwise excitement ground-truth data allows us to study the effectiveness of each of our five cues in predicting how exciting a basket is. For our evaluations, we ignore the baskets which were hard to decide and focus only those baskets which had atleast 2/3rd agreement among the users, i.e. atleast 10 out of 15 must agree that one basket is more exciting that the other. This lets us study the effectiveness of our cues in the absence of noise from the hard-to-decide baskets. From Table \ref{tab:inter-rater-table}, we can see that this gives us 625 baskets for evaluations.

For this evaluation, each of the 625 A/B pairs that was shown to the users, is given as input to our system. For each individual cue, the system decides which basket is more exciting. The output of our system for each cue is then compared against the majority decision made by the users for that basket. If the system decision is same as the user decision, then we have a match. For each cue, we also compute the Matthews Correlation Coefficient (MCC) that gives the amount of agreement or disagreement between the system decision and the user decision \cite{matthews1975comparison, powers2011evaluation}. If MCC is -1, it means that the system decision and the user decision are in total disagreement. If MCC is +1, it means that the system decision and the user decision are in total agreement. If the MCC is 0, it means that the match is decision is no better than random.

The performance of each individual cue is shown in figure \ref{fig:cue-performance-fig}. We can see that ``Motion", ``Basket Type" and ``Audio" are relatively strong indicators of how exciting a basket is while ``Score Differential" and ``Player Ranking" are very close to being no better than random. Out of the five cues, ``Audio" is the strongest indicator with a MCC score of 0.371.

\subsection{Evaluating Weighted Cue Combination}

To learn the weights of the various cues, we perform 25-fold cross-validation where we hold out all the baskets from 1 game for testing while using all the baskets from the other 24 games for learning the weights. The process is repeated 25 times, with a different game being held out for testing in each run. In each run, all combinations of weights are tried and the combination that results in the most matches with the user decision on the held out test game is deemed as the winning set of weights for that run. After all the runs are complete, we average the weights across the 25 runs to get the final set of weights.

\begin{figure}
\begin{centering}
\includegraphics[width=0.8\columnwidth]{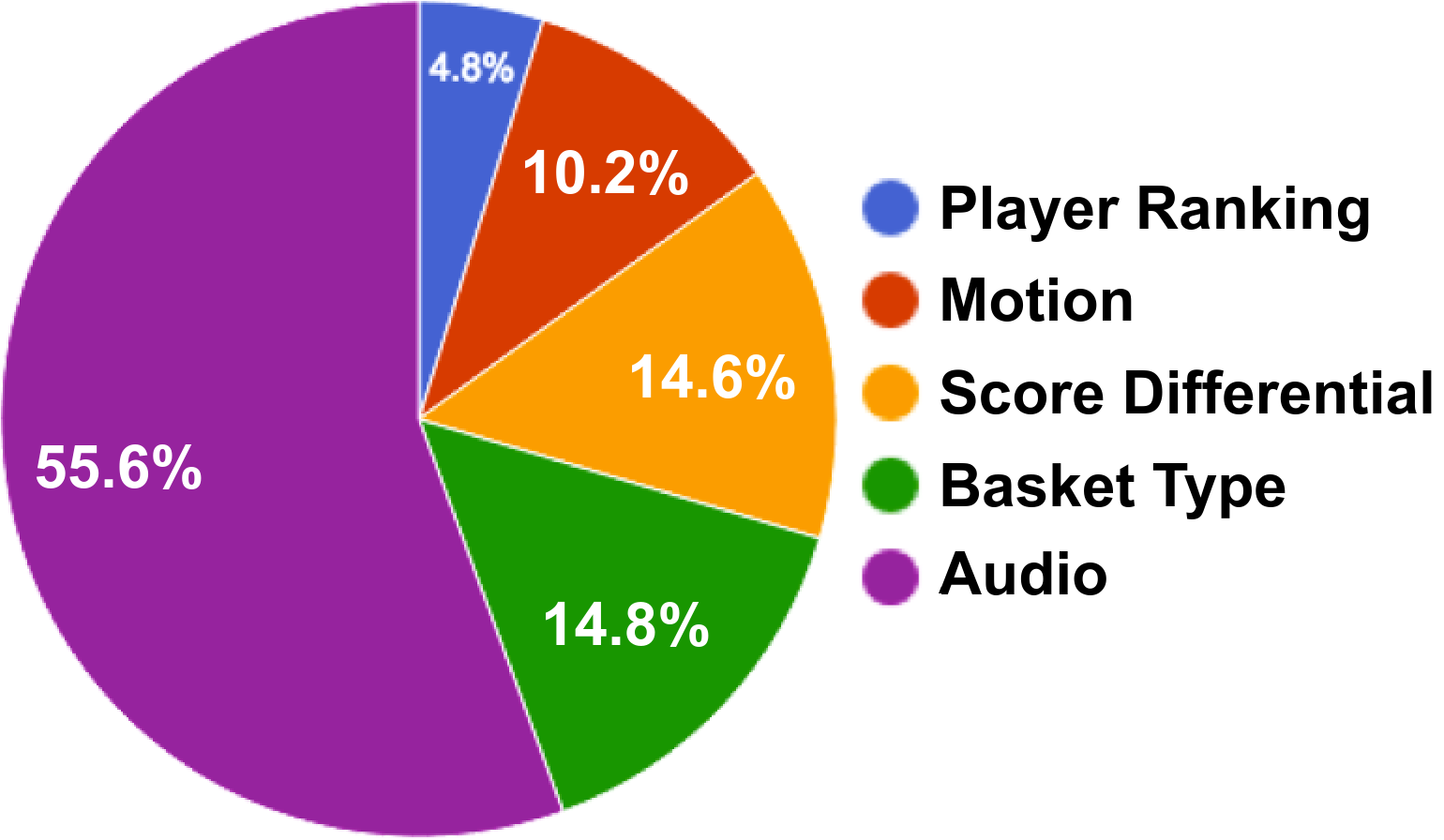}
\par\end{centering}
\caption[Weight of each individual cue in the final cue combination]{\label{fig:cue-weights}The percentage of the total weight that each cue gets during weighted cue combination (after running 25-fold cross-validation).}
\end{figure}

The average percentage of baskets that matched user decision across 25 runs was 75.33\%. The lowest average percentage was 52.81\% with the highest was 91.90\%. When the weights are averaged across all the 25 runs and normalized to add upto one, as shown in figure \ref{fig:cue-weights}, we see that \textbf{Player Ranking} gets \textbf{4.8\%} of the total weight, while \textbf{Motion} gets \textbf{10.2\%}, \textbf{Score Differential} gets \textbf{14.6\%}, \textbf{Basket Type} gets \textbf{14.8\%}, and Audio gets \textbf{55.6\%} of the total weight respectively. 

It is interesting to see that after combining all the cues, Audio makes up 55.6\% of the total share. This shows that the audience cheers and the loudness and pitch of the commentator does indeed drive the excitement levels during a basketball game. It is also interesting to note that the top three dominant cues, Audio, Basket Type, and Score Differential, which make up 85\% of the total weight are all contextual cues derived from the environment. This highlights the importance of the use of contextual cues in our approach.

\begin{table}[]

\caption[Statistical significance of cue combination over each cue]{\label{tab:cue-comb-stat-sig} McNemar's tests on statistical significance between each individual cue vs. cue combination.}

\centering
\begin{tabular}{|l|r|r|}
\hline
                                       & \multicolumn{1}{c|}{\textbf{$\chi^2$}} & \textbf{$p$-value}          \\
                                       \hline
                                       \hline
Player Ranking vs. Cue Combination     & 86.98                   & \textless 0.0001 \\ \hline
Score Differential vs. Cue Combination & 63.00                   & \textless 0.0001 \\ \hline
Motion vs. Cue Combination             & 31.74                   & \textless 0.0001 \\ \hline
Basket Type vs. Cue Combination        & 57.34                   & \textless 0.0001 \\ \hline
Audio vs. Cue Combination              & 32.48                   & \textless 0.0001            \\ \hline
\end{tabular}

\end{table}

When the final average weights are used in cue combination and the system output is compared with the user decision, the percentage of baskets that match the user decision is \textbf{76.4\%}. This is a significant improvement over the best percentage obtained by any single cue (62.6\% with Audio, see \ref{fig:cue-performance-fig}). Furthermore, the MCC score goes up to \textbf{0.528} (from the previous best of 0.371 with only Audio). In order to ensure that the improvement we see from cue combination over each of the individual cues is statistically significant, we ran the McNemar's chi-square test (with Yates' continuity correction). The $null$ hypothesis says that the improvements we see after cue combination are due to chance., However as shown in Table \ref{tab:cue-comb-stat-sig}, the $\chi^2$ values are greater than the critical value (at 95\% significance level) of 3.84 and the $p$-values are less than the significance level ($\alpha$) of 0.05. Thus the $null$ hypothesis can be rejected allowing us to conclude that the improvements seen after cue combination are \emph{not} due to chance.

\subsection{Evaluating Highlights}

With the final weights for each cue, we can now rank order all the baskets in a game by their excitement score, pick the top $N$ baskets, order them by the game clock (so that they are in chronological order), and generate the highlight video. For the 10 games for which we have ESPN highlights to compare against, we pick $N$ as the number of baskets that ESPN put in their highlights for each of those 10 games. For the other 15 games, we pick $N$ to be 10 (the average number of baskets that ESPN shows in their highlights).

\begin{table}[]

\caption[A/B testing cue combination highlights vs. individual cue highlights with all users]{\label{tab:ab-cue-individual}A/B test results with \textbf{all users}: Cue combination highlights vs highlights generated using the individual cues.}

\centering
\begin{tabular}{|l|r|r|}
\hline
\multicolumn{1}{|c|}{\begin{tabular}[c]{@{}c@{}}\textbf{Cue Combination}\\ \textbf{vs.}\end{tabular}} & \multicolumn{1}{c|}{\begin{tabular}[c]{@{}c@{}}\textbf{Num games}\\ \textbf{for which cue}\\ \textbf{combination}\\ \textbf{highlights}\\ \textbf{was selected}\\ \textbf{by user majority}\end{tabular}} & \multicolumn{1}{c|}{\begin{tabular}[c]{@{}c@{}}\textbf{Median}\\ \textbf{user}\\ \textbf{agreement}\\ \textbf{percentage}\end{tabular}} \\
\hline
\hline
Player Ranking      & 22 / 25                                                                                                                          & 61.29\%                                                                      \\ \hline
Score Differential  & 17 / 25                                                                                                                          & 54.84\%                                                                      \\ \hline
Motion              & 17 / 25                                                                                                                          & 58.06\%                                                                      \\ \hline
Basket Type         & 16 / 25                                                                                                                          & 58.06\%                                                                      \\ \hline
Audio               & 16 / 25                                                                                                                          & 61.29\%                                                                      \\ \hline
\end{tabular}

\end{table}

\begin{table}[]

\caption[A/B testing cue combination highlights vs. individual cue highlights with only basketall fans]{\label{tab:ab-cue-individual-fans}A/B test results with \textbf{basketball fans}: Cue combination highlights vs highlights generated using the individual cues.}

\centering
\begin{tabular}{|l|r|r|}
\hline
\multicolumn{1}{|c|}{\begin{tabular}[c]{@{}c@{}}\textbf{Cue Combination}\\ \textbf{vs.}\end{tabular}} & \multicolumn{1}{c|}{\begin{tabular}[c]{@{}c@{}}\textbf{Num games}\\ \textbf{for which cue}\\ \textbf{combination}\\ \textbf{highlights}\\ \textbf{was selected}\\ \textbf{by user majority}\end{tabular}} & \multicolumn{1}{c|}{\begin{tabular}[c]{@{}c@{}}\textbf{Median}\\ \textbf{user}\\ \textbf{agreement}\\ \textbf{percentage}\end{tabular}} \\
\hline
\hline
Player Ranking                            & 20 / 25                                                                                                                                               & 60.00\%                                                                                           \\ \hline
Score Differential                        & 14 / 25                                                                                                                                               & 60.00\%                                                                                           \\ \hline
Motion                                    & 18 / 25                                                                                                                                               & 60.00\%                                                                                           \\ \hline
Basket Type                               & 16 / 25                                                                                                                                               & 60.00\%                                                                                           \\ \hline
Audio                                     & 15 / 25                                                                                                                                               & 60.00\%                                                                                           \\ \hline
\end{tabular}

\end{table}

\begin{figure*}
\begin{centering}
\includegraphics[width=1.0\textwidth]{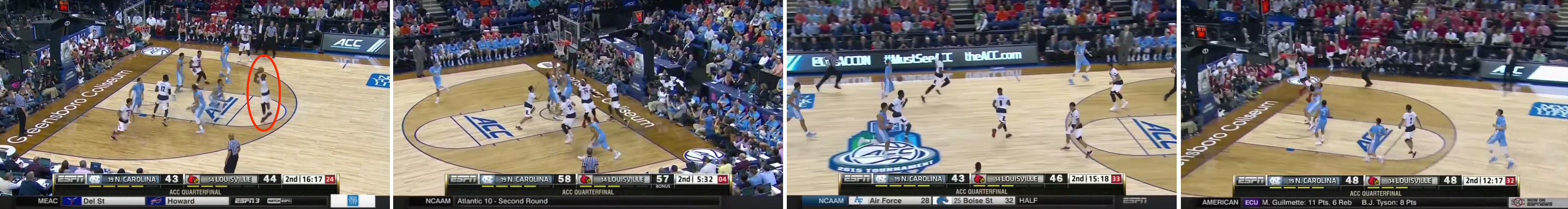}
\par\end{centering}
\caption[Sample frames from highlights generated using individual cues]{\label{fig:cue-each}Sample frames from highlights generated using individual cues. From left to right: Player Ranking, Score Differential, Motion, and Basket Type.}
\end{figure*}

\textbf{Cue Combination vs. Individual Cues:} We generated highlights for all the 25 games using cue combination and also using each of the five individual cues. This gave us 6 highlight videos for each game. Figure \ref{fig:cue-each} shows 4 sample frames from the highlights generated for the Louisville vs. North Carolina NCAA game. Highlights generated using only Player Ranking mostly featured baskets by Terry Rozier (circled in red) who was Louisville Cardinals' star player (now a NBA draft pick for the Boston Celtics). The highlights generated using only Score Differential featured ``neck-to-neck" baskets (for example, a score of 58-57 can be seen in the figure) while the highlights generated using only Motion featured baskets with lots of player and camera motion (as seen in the figure). Finally, the highlights generated using only Basket Type featured mostly Dunk shots (a sample shot is shown in the figure).

In order to see if users prefer cue combination highlights over the highlights generated by individual cues, we ran another A/B test user study. Given the harder nature of this study where users have to watch two minute-long clips, we doubled the number of users per test from 15 to 31. For each game, the users were asked to pick the highlight that they preferred. Similar to the previous study, sufficient care was taken to randomize the A/B pairs and the users were asked to fill in a similar questionnaire mentioning their age, if they are a basketball fan, if they are a fan of one of the teams, and the number of games they watched last season.

This study had 335 unique users out of which 245 are basketball fans. The results of the study with all 335 users are shown in Table \ref{tab:ab-cue-individual} and the results with only the 245 basketball fans is shown in Table \ref{tab:ab-cue-individual-fans}. We can see that highlights generated using cue combination were preferred over the highlights generated using the individual cues by both regular users and basketball fans. However, it is interesting to note that basketball fans seem to prefer score differential highlights slightly more than regular users. This could be due to the fact that basketball fans watch the game more closely and pay more attention to the scores shown on the graphics overlay.

\textbf{Cue Combination vs. ESPN:} Our dataset contains the highlights produced by ESPN for 10 out of the 25 games. We ran a similar A/B test study with 31 users where users were shown the ESPN highlight and our cue combination highlight and were asked to pick the highlight that they preferred. To make the comparison fair, we regenerated the ESPN highlights using the same video production pipeline that we used to produce our highlights. This ensured that both the highlights shown in the A/B tests are visually similar.

The results of the A/B tests showed that among all users, our highlights were preferred in 5/10 games and the ESPN-ranked highlights were preferred in the other \textbf{5/10} games. The median agreement percentage was 51.61\%. Among basketball fans, our highlights were preferred in \textbf{7/10} games while the ESPN-ranked highlights were preferred in 3/10 games. The median agreement percentage was 53.33\%. Although basketball fans showed a slight preference to our highlights, the median agreement percentage shows that the decision was really hard to make. This shows that the users had a tough time picking between our highlights and ESPN-ranked highlights which in-turn indicates that we are performing as well as ESPN in picking baskets for producing basketball highlights.

\begin{figure}
\begin{centering}
\includegraphics[width=1.0\columnwidth]{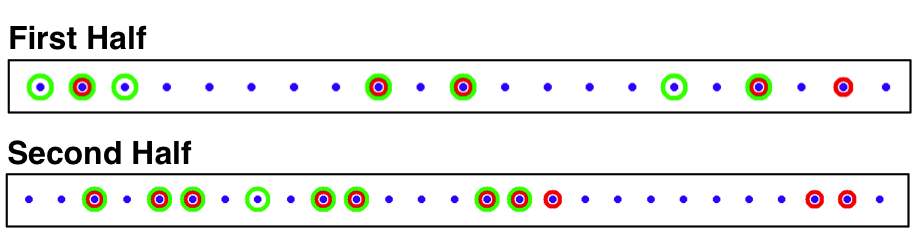}
\par\end{centering}
\caption[Common baskets picks across out highlights and ESPN highlights]{\label{fig:picks}Basket picks for a sample game (Duke vs. Florida State, 9th Feb 2015). Each blue dot represents a basket that occurred during the game play. Baskets with red circles were picked by ESPN for their highlights and baskets with green circles were picked by our method for our highlights. We can see that 11 out of the 15 baskets were commonly picked (4 overlaps in the first half and 7 overlaps in the second half).}
\end{figure}

Comparing the individual baskets that were picked for the highlights across all the 10 games, we noticed 67.4\% overlap in the baskets that we picked and the baskets that ESPN picked. This is illustrated in figure \ref{fig:picks} where we show the basket picks for a sample game (Duke vs. Florida State, 9th Feb 2015). We can see that across both the game periods, out of the 15 baskets, 11 baskets were commonly picked by our cue-combination approach and by ESPN. The probability of these 11 baskets being picked in common, by chance, is 0.00005.

\begin{figure}
\begin{centering}
\includegraphics[width=1.0\columnwidth]{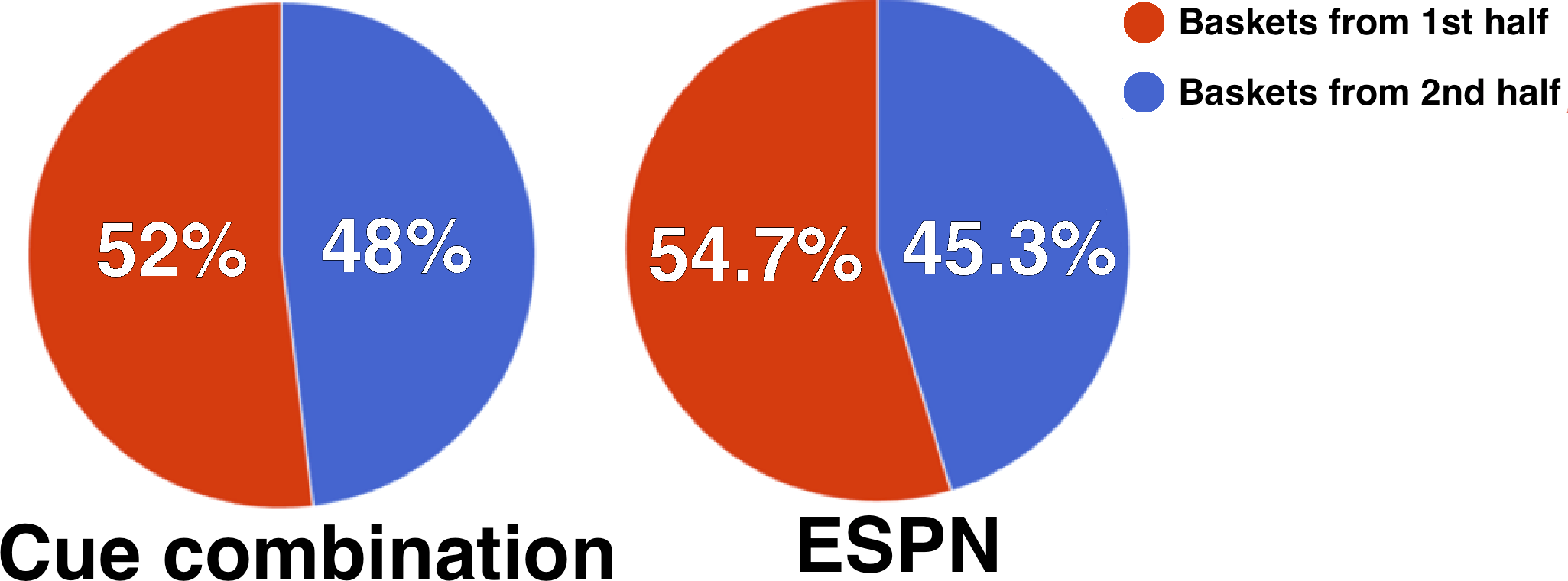}
\par\end{centering}
\caption[Distribution of the baskets across the two halves of the game]{\label{fig:halves}The distribution of baskets across the two halves of the game for 10 games. \textbf{Left:} The basket distribution for our cue combination highlights. \textbf{Right:} The basket distribution for ESPN highlights. We can see that our approach isn't biased towards any particular half of the game and closely follows ESPN's distribution across the two halves.}
\end{figure}

Another factor to consider is the distribution of the baskets shown in the highlights across the two halves of the game. As shown in figure \ref{fig:halves}, for the 10 games, our cue combination picks 48\% of the baskets from the first half of the game and 52\% of the baskets from the second half of the game. Looking at the ESPN highlights for those 10 games, we can see that ESPN has a very similar distribution. They pick 45.3\% of the baskets from the first half and 54.7\% of the games from the second half. This shows that our approach isn't biased towards baskets from any single period of the game and closely follows ESPN's distribution.

Figure \ref{fig:picks} and figure \ref{fig:halves} further highlights the practicality of our approach and the similarity of our highlights to the ESPN-ranked highlights. Also, this gives us an insight into why the users had a hard time deciding between our highlights and ESPN-ranked highlights in the A/B test user studies.

\section{Discussion}

Our comprehensive user studies and analysis have shown the effectiveness of leveraging contextual cues derived from the environment in determining the exciting baskets of a game and automatically producing the game highlights. 

However, there are some limitations of our work. The ground-truth data that we used for evaluations is the pairwise excitement gathered using our A/B tests. Pairwise excitement scores are inherently different from true excitement scores wherein the users look at all the baskets in a game and rank orders them from least exciting to most exciting. While this kind of data would be more useful than pairwise excitement scores, it is practically impossible to get the true excitement scores through user studies. A typical basketball game has 50 or more baskets and it is very hard for humans to look at so many baskets, remember their excitement levels, and rank order them. Another limitation is that our method does not adaptively change the weights of the cues based on changes in the environment. For example, if the audience start leaving after the first half, we should be able to detect this and adaptively decrease the weight given to the audio cue. This is a good direction for our future work.

While running the A/B tests where users are shown two basket clips and asked to pick the one that is more exciting, we wondered how the results would change if users were shown the same A/B pair, but with only the audio and only the video. To answer this question, we picked the top 203 baskets that had 80\% (12/15) agreement between the users (i.e. baskets for which most users had a strong consensus in deciding between choice A and choice B), and generated two more A/B tests with the same pairs as before. But this time one of the A/B tests had just the audio of the game and the other A/B test had clips with just the video (with the audio stripped out). We ran these new A/B tests for the 203 baskets with 15 users per test (as before) and analyzed the results. It is interesting to note that for the audio-only case, the agreement fell from 80\% to \textbf{59.70\%} and the Fleiss' kappa value dropped from 0.270 (fair agreement) to \textbf{0.177} (slight agreement), and for the video-only case, the average pairwise agreement fell to \textbf{59.82\%} and the Fleiss' kappa value dropped to \textbf{0.116} (slight agreement). This shows that the strong consensus between the users broke down when they had to pick between clips that had either only the audio or only the video.

Another interesting question is ``how would a highlight generated using randomly selected baskets compare against our cue combination highlights and against ESPN-ranked highlights?". To answer this question, we generated 10 highlights using random baskets selected from 10 of the games (and placed in chronological order) and ran additional A/B tests for evaluating these random highlights against our cue combination highlights and also against the ESPN-ranked highlights. As before, we ran the tests with 31 users per A/B test. The results showed that among all users our cue-combination highlights were preferred over random highlights in \textbf{7/10} games with \textbf{61.29\%} median user agreement. Basketball fans preferred our highlights in \textbf{8/10} games with \textbf{53.33\%} median user agreement. The A/B test results against ESPN-ranked highlights showed that among all users ESPN highlights were preferred over random highlights in \textbf{7/10} games with a median user agreement of \textbf{54.84\%}. Basketball fans preferred ESPN-ranked highlights in \textbf{9/10} games with a median user agreement of \textbf{60.00\%}. These results tell us that while users prefer our highlights and ESPN-ranked highlights over random highlights, the random highlights still do get picked in some of the games. This further highlights the subjective and complex nature of the problem domain.

\section{Conclusion}

In this paper, we explored the use of context derived from the environment along with the visual cues to automatically produce basketball highlights. We explored 5 cues that are indicative of excitement levels in basketball games. We introduced a new dataset of 25 NCAA games with 1,173 baskets with ground-truth pair-wise excitement scores for evaluating our approach. We conducted comprehensive user studies with multiple participants which showed the effectiveness of our cues and our cue combination method that can produce highlights that are comparable to those produced by ESPN. Interesting directions for future work include exploring methods to collect more ground-truth excitement data in a way that maximizes inter-rater reliability, exploring more cues that are indicative of excitement levels, and dynamically adapting the weights of the cues as the game proceeds based on the changes in the environment.



%
\bibliographystyle{abbrv}
\bibliography{sports}  
%
%

\end{document}